\documentstyle[epsfig]{aipproc}
\begin{document}
\setlength{\parskip}{0pt}
\title{Kilo-Hertz QPO and X-ray Bursts\\
in 4U 1608-52 in Low Intensity State }
\author{W. Yu$^{\dagger}$, S. N. Zhang$^*$$^{\star}$, B. A.
Harmon$^*$, W. S. Paciesas$^*$$^{\ddagger}$, C. R. Robinson$^*$$^{\star}$, 
J. E. Grindlay$^{\diamond}$, P. Bloser$^{\diamond}$, D.
Barret$^{\triangleright}$, 
E. C. Ford$^{\sharp}$, M. Tavani$^{\sharp}$, P. Kaaret$^{\sharp}$}
\address{$^*$Marshall Space Flight Center, Huntsville, AL 35810\\
$^{\star}$Universities Space Research Association\\
$^{\dagger}$LCRHEA, Institute of High Energy Physics, Beijing, 100039, China\\
$^{\ddagger}$University of Alabama in Huntsville, Huntsville, AL 35889\\
$^{\diamond}$Harvard-Smithsonian Center for Astrophysics, 60 Garden Street,
Cambridge, MA 02138\\
$^{\triangleright}$Centre d'Etude Spatiale des Rayonnements, CNRS-UPS, France\\
$^{\sharp}$Columbia Astrophysics Laboratory, Columbia University, New York, NY 10027\\}
\maketitle
\begin{abstract}
We present the results from RXTE/PCA observations of 4U 1608-52 in its island 
state on March 15, 18 and 22 of 1996. Three type I X-ray bursts were detected 
in one RXTE orbit on March 22. We observed QPO features peaking at 567-800 Hz 
on March 15 and 22, with source fractional rms amplitude of 13\%-17\% and 
widths of 78-180 Hz in the power density spectra averaged over each spacecraft 
orbit. The rms amplitudes of these QPOs are positively correlated with the 
photon energy. The three X-ray bursts, with burst intervals of 16 and 8 
minutes, have a duration of 16s. The blackbody emission region of the smallest 
X-ray burst among the three suggest it was a local nuclear burning. We also 
discuss a type I X-ray burst candidate in the observation.
\end{abstract}
\section*{Introduction}
4U 1608-52 is a transient X-ray burster with recurrence intervals from 80 days to about 
2 years \cite{lewin93}\cite{lochner94}. It was classified as an atoll
source based on the correlated X-ray spectral variability and High-Frequency-
Noise (HFN) \cite{hk89}\cite{yoshida93}. 4U 1608-52 is known to show spectral and
timing characteristics similar to those in blackhole candidates (BHCs) in the 
low state\cite{klis94}. When in a low luminosity state ($L_{x}\leq{10}^{37}$erg/s), its 
energy spectrum above 2 keV is usually dominated by a power-law
component\cite{yoshida93}\cite{mitsuda89}\cite{penninx89}, which is gradually
cut off above $\sim60-70$ keV\cite{zhangsn96}.

An outburst was observed from 4U 1608-52 with RXTE in early 1996 \cite{marshall96}. 
A few RXTE/PCA pointings on March 3, 6, 9 and 12 show kilo-hertz QPO at 
690-890 Hz \cite{berger96}. Here we present our analysis results obtained
from RXTE/PCA observations on March 15, 18 and 22 with source luminosity
$\sim{10}^{36}$ erg/s. 
\section*{Observations and Data Analysis}
The X-ray monitoring with RXTE/ASM shows that our observations on March 15, 18
and 22 were taken near the end of the outburst when source fluxes were 
below 50 mCrab. The persistent X-ray flux 
(2-20 keV) was $(4.6-11)\times{10}^{-10}$erg/s/cm${}^{2}$. The count rate 
ranged between 190-450 cps (1-60 keV) with source intensity below previous RXTE 
observations \cite{berger96}. Three X-ray bursts were observed in one orbit on
March 22.
\begin{figure*}[t]
\begin{minipage}[t]{6.5cm}
\psfig{file=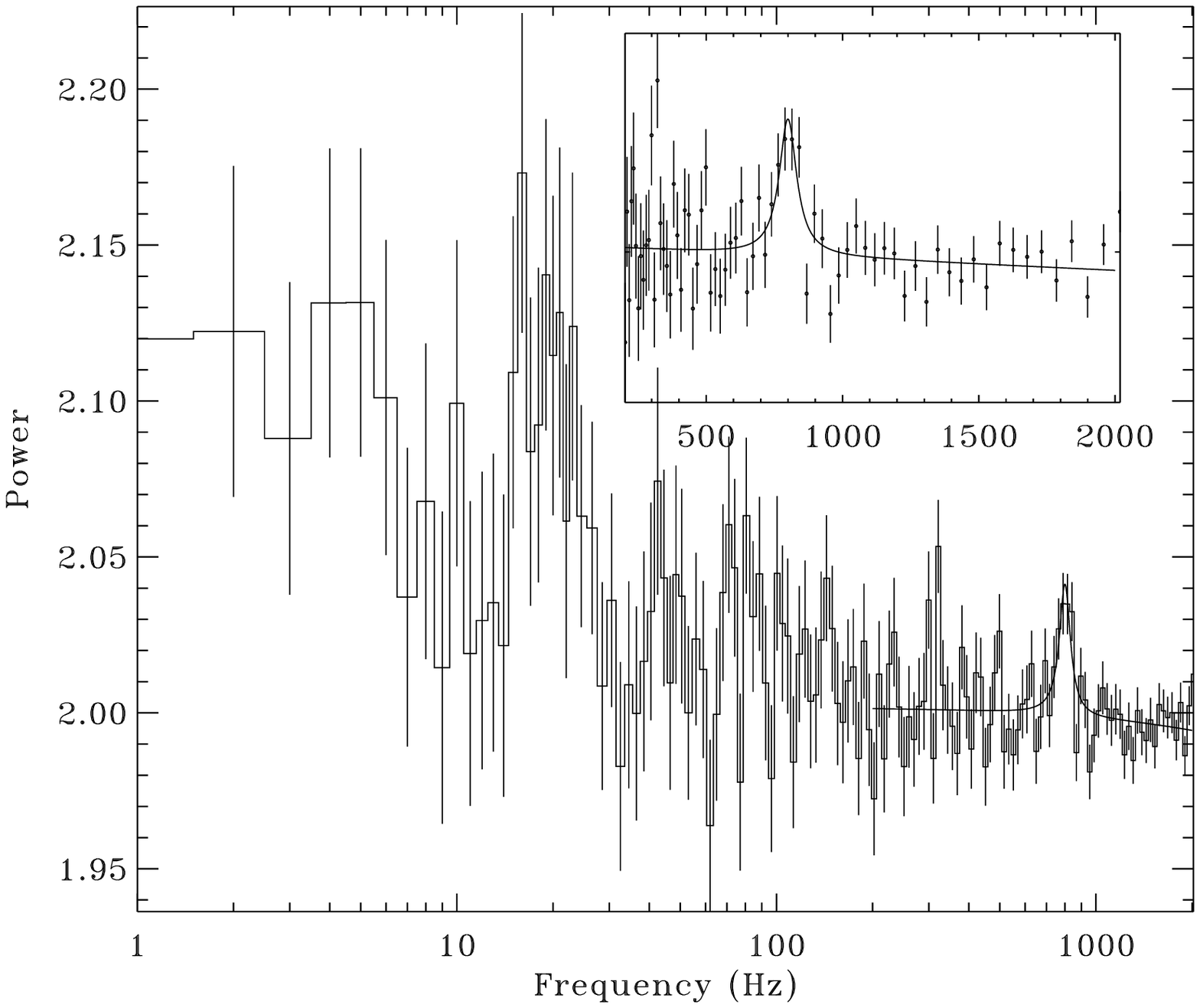,width=6.5cm}
\caption{\scriptsize The QPO peak at 800 Hz in the first orbit on March 15. The PDS is 
complex and a broad peak at 20 Hz is also visible.}
\end{minipage}
\hspace{0.05cm}
\begin{minipage}[t]{6.5cm}
\psfig{file=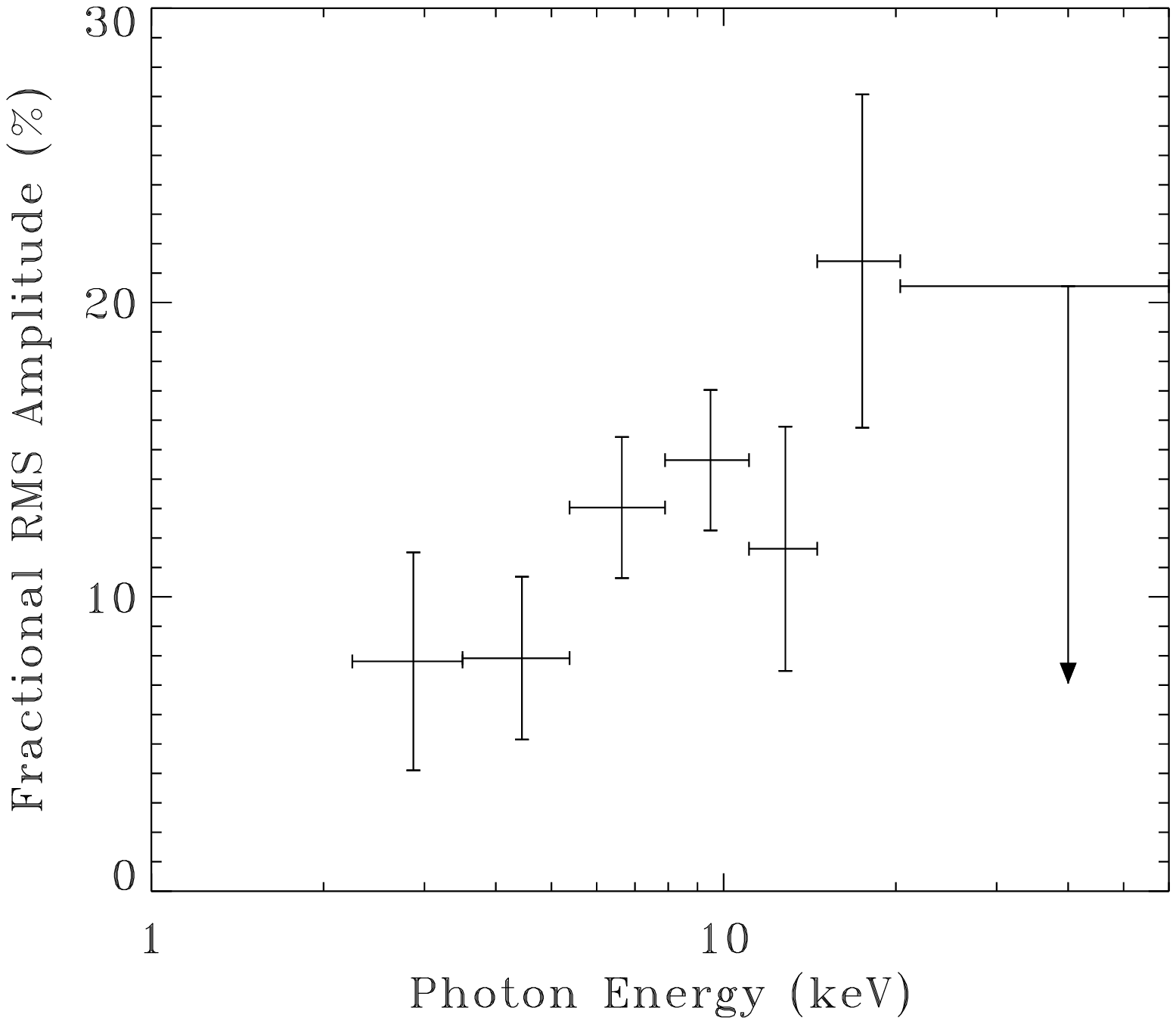,width=6.5cm}
\caption{\scriptsize Average fractional {\it rms} amplitude of QPO as a function 
of photon energy. We average the results obtained from {\it Event Mode} data 
analysis of the 6 orbits.}
\end{minipage}
\hspace{0.05cm}
\end{figure*}
\subsection*{Power Density Spectra (PDS)}
By analyzing the {\it Event Mode} data with background subtraction, we have 
found that the Power Density Spectra (PDSs) were dominated by HFN with rms 
10\%-14\% in our observations. We detect kilo-hertz QPO features at 
567-800 Hz in 6 orbits 
, with source fractional rms amplitude of 13\%-17\% and widths of 78-180 Hz. 
As an example, Fig. 1 is the PDS obtained from the 1-30 keV {\it Event Mode} 
data averaged over the first orbit on March 15. 
We also investigate the energy dependence of the QPO features. 
A positive correlation between the average QPO rms amplitude and the photon 
energy is observed, as shown in Fig. 2. No kilo-hertz pulsations have been 
detected at 99\% confidence level in the X-ray bursts in our preliminary 
searching. The details of the timing analysis will be reported elsewhere 
\cite{yu97}. 
\begin{figure*}[t]
\begin{minipage}[b]{6.5cm}
\psfig{file=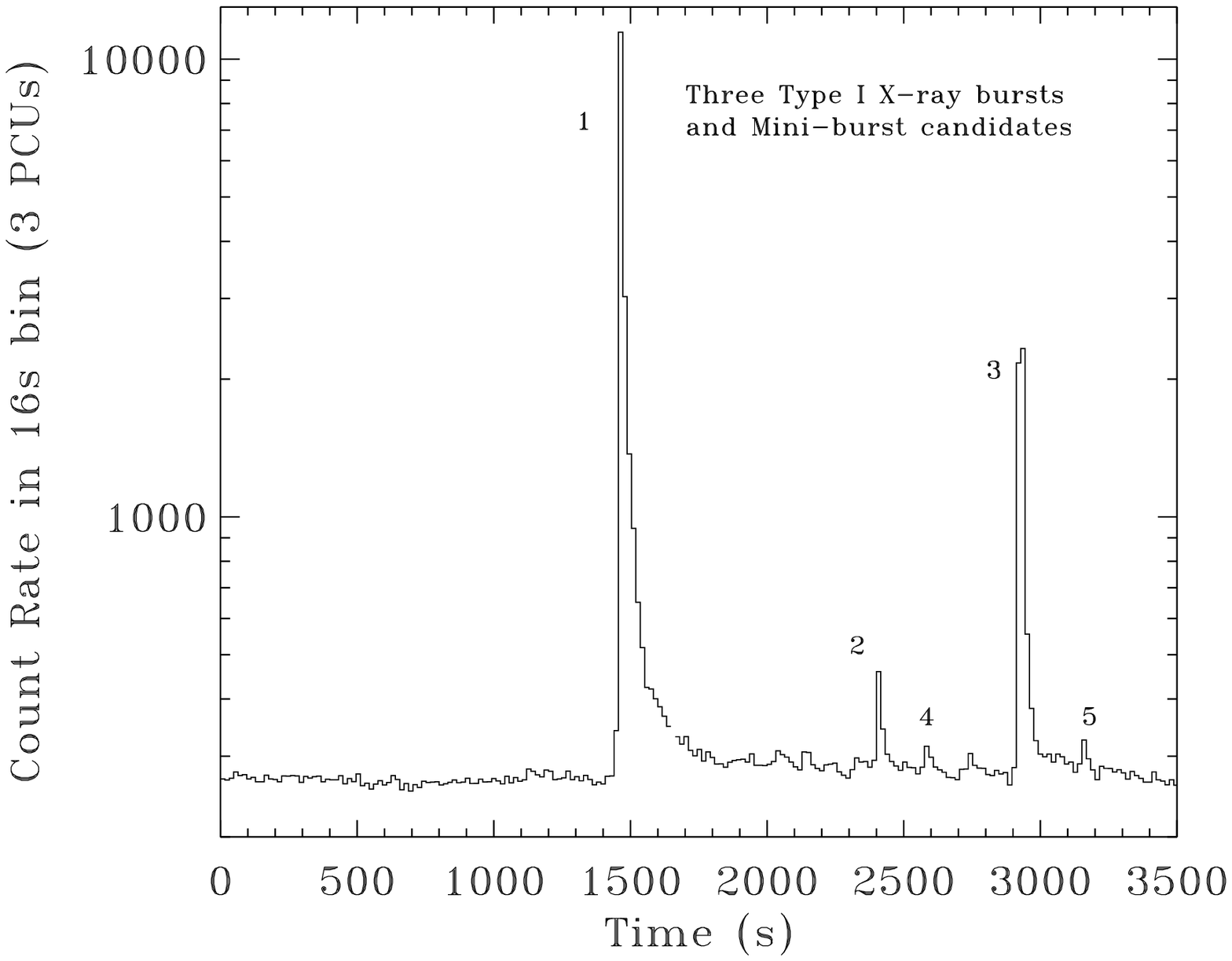,width=6.5cm}
\caption{\scriptsize Three type I X-ray bursts (marked as 1-3) and mini-burst 
candidates (marked as 4-5). The time resolution is 16s and the count rates 
were obtained from 3 PCUs.}
\end{minipage}
\hspace{0.05cm}
\begin{minipage}[b]{6.5cm}
\psfig{file=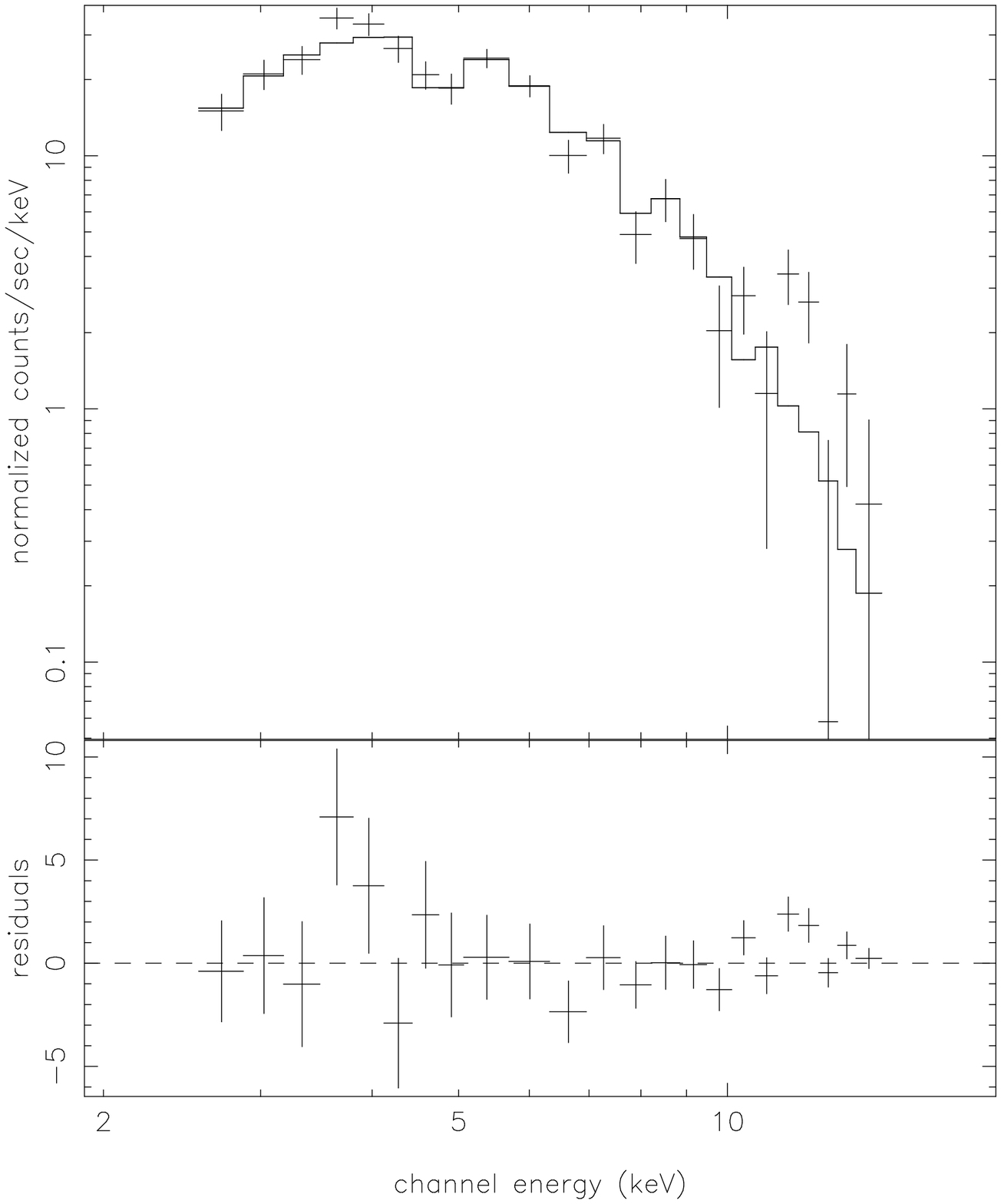,width=6.5cm}
\caption{\scriptsize Energy spectrum of burst 2 with $kT_{BB}=1.2$ keV and
${\chi}^{2}/21(dof)=1.49$. }
\end{minipage}
\hspace{0.05cm}
\end{figure*}
\begin{figure*}[b]
\begin{minipage}[b]{6.5cm}
\psfig{file=f5.ps,width=6.5cm}
\caption{\scriptsize Time profile of one of the burst candidates (5). A model composed of
a linear rise and an exponential decay was fit to the data.
 }
\end{minipage}
\hspace{0.05cm}
\begin{minipage}[b]{6.5cm}
\psfig{file=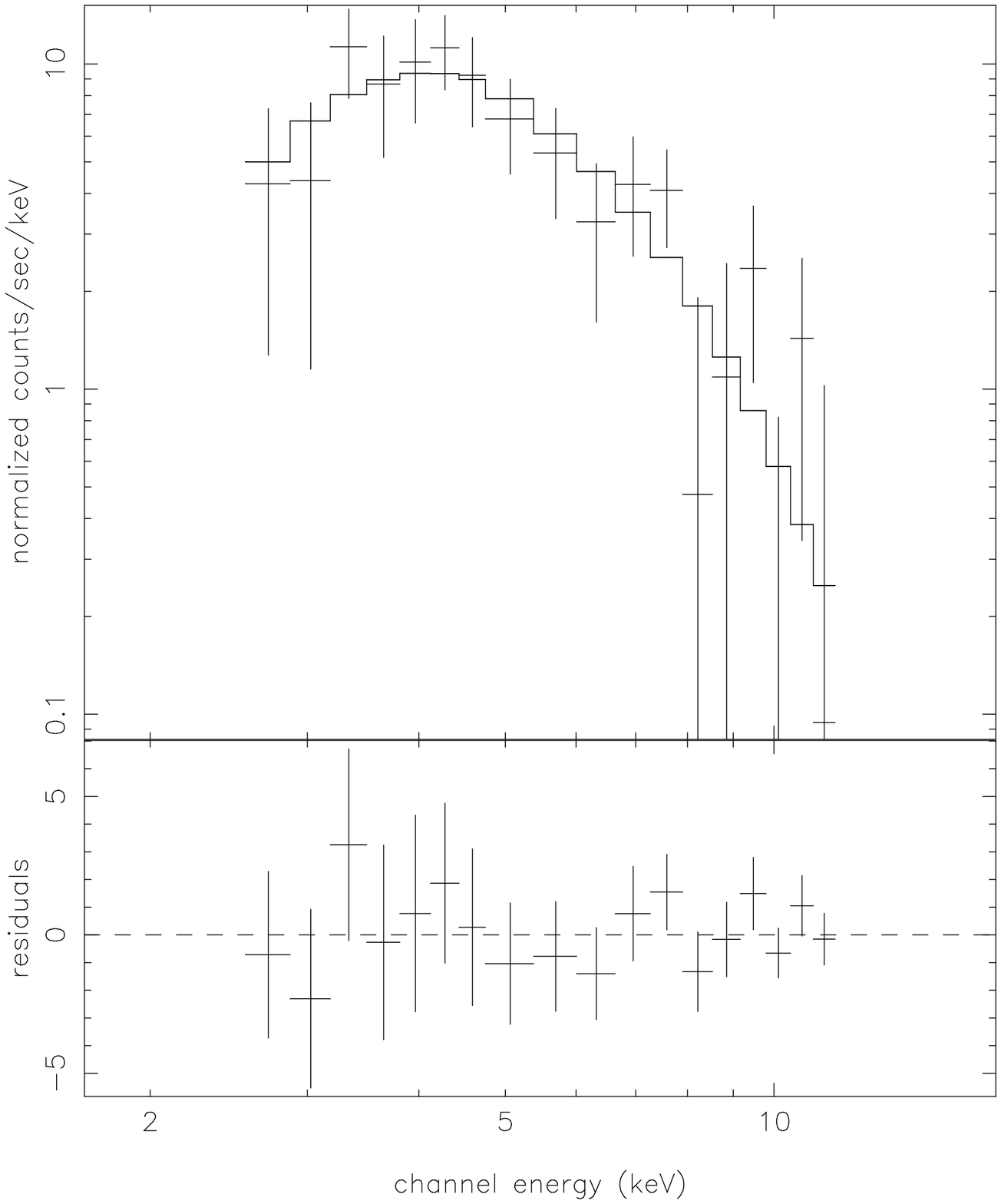,width=6.5cm}
\caption{\scriptsize The Energy spectrum of the burst candidate (5) with $kT_{BB}=1.2$ keV
and ${\chi}^{2}/16(dof)=0.51$.}
\end{minipage}
\hspace{0.05cm}
\end{figure*}
\subsection*{X-ray Bursts and Burst-like Variations}
Three X-ray bursts were observed in the second orbit on March 22, 1996. They
all have profiles consistent with a fast rise and exponential decay, and show 
blackbody-type energy spectra, indicating that they are type I 
X-ray bursts. Fig. 3 is the light curve with these bursts. The lightcurve is 
obtained from 3 PCUs and with 16 s resolution. The properties of these bursts 
are listed in Table 1. The average BB radii were obtained from spectral 
fitting to a BB model with $N_{H}=1.5\times{10}^{22}{cm}^{-2}$.
\begin{table}
\caption{Properties of Type I X-ray Bursts}
\label{table1}
\begin{tabular}{lcccc}
\multicolumn{1}{c}{Burst No.}&{Peak Count Rate (cps)\tablenote{from 3 PCUs of
RXTE/PCA}}&{
Duration (s)\tablenote{defined as the ratio between the integrated burst 
counts and the burst peak count rate in PCA band}}&{BB Radius (km)\tablenote{from results of spectral
fitting to a blackbody(BB) model with a correction assuming 
that the ratio between the color temperature and the effective temperature 
is 1.5}}\\
\tableline
{  1}&16500&$16.7\pm0.1$&$\sim$ 13 \\
{  2}&240&$15.4\pm1.0$&$\sim$ 4 \\
{  3}&4150&$16.7\pm0.3$&$\sim$ 18 \\
\end{tabular}
\end{table}

The burst durations indicate that they belong to the "slow" class of bursts 
usually observed in 4U 1608-52 in its low state 
\cite{murakami80}. The intervals between the bursts were about 16 min and 8.5 min,
among the shortest now known \cite{lewin93}. 
The ratio between the average persistent 
flux and the average burst flux of burst 3, regarding the relative time
interval of 24 minutes between burst 1 and 3, was smaller than 6. 
In Fig. 4, we show the average energy spectrum of the burst 2 with the best fit 
BB model with $N_{H}=1.5\times{10}^{22}{cm}^{-2}$. The burst BB 
temperature was $1.2\pm0.3$ keV. The derived BB emission radius is smaller 
than the radii obtained for burst 1 and 3 ( Table 1.). 

Marked as 4 and 5 in Fig. 3 were 
burst-like intensity variations which were more than $4\sigma$ above the 
persistent intensity level in the light curve with the 10 s resolution during the
entire observations. 
The second one was stronger. Its peak count rate in 3 PCUs was $90\pm18$ cps. 
Its profile can be fitted to a model with a linear rise and an exponential 
decay (see Fig.5). The derived rise time (from 10\% to 90\% of 
peak flux) and decay time (e-folding time) are $7.6\pm1.3$ s and 
$12.9\pm2.5$ s respectively. For the burst 2, the rise time and decay time 
are $8.2\pm1.2$ s and 
$13.1\pm1.5$ s. So they have similar profiles. In addition, the spectrum of this burst-like
variation is consistent with the BB model, but a power law model can not be 
excluded. Fitting its spectra to a BB model yield a blackbody temperature of $1.2\pm0.1$ 
keV (Fig.6). Its corrected radius of BB emission region is 2.2 km. 
The hardness ratio (5.4-20.4 keV/2.2-5.4 keV) during the decay was also 
similar to that in burst 2.
\section*{Discussion and Conclusion}
Strong HFN components in PDSs and low X-ray luminosity (2-20 keV) of 
$(0.7-1.7)\times{10}^{36}$erg/s assuming a distance of 3.6 kpc suggest 4U
1608-52 was in its island state \cite{hk89}. 
The QPO features in our observations show a similar relation
between QPO rms amplitude and photon energy to those observed in a higher 
intensity state \cite{berger96}. 
The burst 3 cannot be explained as being produced by replenishing 
sufficient fuel through accretion within the short interval between burst 3 
and burst 1. This suggest that a portion of the nuclear fuel had survived 
the previous bursts \cite{lewin93}. 
The burst 2 had a BB emitting radius smaller than 
those of the two stronger bursts and the size of the canonical neutron 
star radius. One burst-like intensity variation, which could not be 
observed with EXOSAT Medium Energy (ME) Experiment (estimated EXOSAT ME 
peak count rate (1-20 keV) is 12.4 c/s), also shows signatures of a type 
I X-ray burst with an even smaller BB emitting radius. This strongly 
supports the local nuclear burning interpretation. Our results suggest 
that these small bursts or variations, only observed associated with the 
strong bursts in our observations, may be attributed to the mixing 
mechanism during the post burst phase of the strong bursts \cite{woosley85}.  
\\

We appreciate various assistances from the members of RXTE/GOF. WY thanks 
Prof. R. E. Taam of Northwestern University and Dr. William Zhang of GSFC 
for many helpful suggestions. WY would like to acknowledge 
the support from the National Natural Science Foundation of 
China.
 

\begin{references}
\bibitem{lewin93}Lewin W.H.G, van Paradijs J., and Taam R.E., {\it Space
Sci.Rev.} {\bf 62}, 223 (1993)
\bibitem{lochner94}Lochner J.C., Roussel-Dupr\'{e} D. {\it ApJ} {\bf 435}, 840
(1994)
\bibitem{hk89}Hasinger G., van def Klis M., {\it A\&A} {\bf 225}, 79 (1989)
\bibitem{yoshida93}Yoshida K., et al., {\it PASJ } {\bf 45}, 605 (1993)
\bibitem{klis94}van der Klis M. {\it ApJS} {\bf 92}, 511 (1994)
\bibitem{mitsuda89}Mitsuda K., et al. {\it PASJ } {\bf 36}, 741 (1989)
\bibitem{penninx89}Penninx W., Damen E., et al. {\it A\&A} {\bf 208}, 146 (1989)
\bibitem{zhangsn96}Zhang S.N., et al. {\it A\&A} {\bf 120}, 279 (1996)
\bibitem{marshall96}Marshall F.E., Angelini L. {\it IAU Circ.}6331 (1996)
\bibitem{berger96}Berger M. et al., {\it ApJ} {\bf 469}, L13 (1996)
\bibitem{yu97}Yu W. et al., submitted to {\it ApJ} (1997)
\bibitem{murakami80}Murakami T., et al. {\it ApJ} {\bf 240}, L143 (1980)
\bibitem{woosley85}Woosley S.E., and Weaver T.A. {\it High Energy Transients 
in Astrophysics, AIP Conf.Proc. 115}, New York: AIP Press, 1985, pp. 273. 
\end{references}
\end{document}